\documentclass[journal,transmag]{IEEEtran}
\usepackage{graphicx}
\usepackage{amsmath,amssymb,amsfonts,amsthm,mathtools,bm}
\usepackage[dvipsnames]{xcolor}
\usepackage{comment}
\usepackage{hyperref}
\usepackage{booktabs,siunitx}

\usepackage{subcaption}

\newcommand{\norm}[1]{\left\lVert#1\right\rVert}

\DeclareMathOperator\supp{supp}
\DeclareMathOperator*{\argmin}{argmin}

\newtheorem{definition}{Definition}
\newtheorem{thm}{Theorem}

\NewDocumentCommand{\B}{}{\fontseries{b}\selectfont}

\usepackage{tikz}
\usepackage{framed}
\colorlet{LightBlue}{blue}

\usetikzlibrary{calc,3d}
\usetikzlibrary{fit,shapes,positioning}
\def\layersep{3.4em}
\def\widthNN{4}

\title{Neural Vector Tomography for Reconstructing a Magnetization Vector Field}

\author{
	\IEEEauthorblockN{Giorgi Butbaia\IEEEauthorrefmark{1},
Jiadong Zang\IEEEauthorrefmark{2}}
\IEEEauthorblockA{Department of Physics and Astronomy,
University of New Hampshire,\\
Durham, NH, 03824, USA\\
Email: \IEEEauthorrefmark{1}giorgi.butbaia@unh.edu,
\IEEEauthorrefmark{2}jiadong.zang@unh.edu}}

\begin{document}
\IEEEtitleabstractindextext{%
\begin{abstract}
Discretized techniques for vector tomographic reconstructions are prone to producing artifacts in the reconstructions. The quality of these reconstructions may further deteriorate as the amount of noise increases. In this work, we instead model the underlying vector fields using smooth neural fields. Owing to the fact that the activation functions in the neural network may be chosen to be smooth and the domain is no longer pixelated, the model results in high-quality reconstructions, even under presence of noise. In the case where we have underlying global continuous symmetry, we find that the neural network substantially improves the accuracy of the reconstruction over the existing techniques.
\end{abstract}

\begin{IEEEkeywords}
vector tomography, reconstruction methods, machine learning
\end{IEEEkeywords}}
\maketitle

\section{Introduction}

Recent years have witnessed the rise of three-dimensional (3D) nanomagnetism \cite{fernandez-pacheco2017}. Besides many exciting breakthroughs in theoretical modeling and experimental synthesis, real space imaging of magnetization distribution in 3D is not yet mature. Existing traditional magnetic imaging tools such as magneto-Kerr effect microscopy or magnetic force microscopy can only reveal the information of magnetization on a surface. X-ray and transmission electron microscopy (TEM) can send the beam through the sample, but conventional technique only get the average magnetic information along the beam direction.

Exciting attempts have been made to develop tomography based on X-ray and TEM \cite{Donnelly_2018,donnelly2020,hierro-rodriguez2020,phatak2010,tanigaki2015,wolf2019,wolf2022}. Beams are sent in different angles and a series of two-dimensional histograms are taken. Central slicing theorem is usually employed to enable 3D imaging retrieval \cite{lai1994,phatak2008}, but its capacity is limited by the missing wedge problem in these 3D magnetic tomography experiments. Instead, iterative optimization methods were used to obtain magnetic structures that minimize the discrepancy between reconstructed and experimental histograms \cite{phatak2015,Donnelly_2018,prabhat2017,Lyu2024}. However, this approach is susceptible to high sensitivity to errors in the measurements. In particular, this may introduce significant artifacts, which can significantly deteriorate the quality of the reconstruction.

Deep learning is another working path towards 3D reconstruction \cite{ruckert2022neat, mildenhall2020nerf,lyu2022}.  
Recent progress in tomographic reconstruction using neural networks has provided an evidence for stability and robustness of the neural-network-based methods under noisy environments \cite{Chang2024, he2020radon, ruckert2022neat, Mishra-Sharma:2022bco, levis2022gravitationally}. In particular, due to the smooth nature of the neural-network approximation of the reconstructed field, the presence of noise does not significantly affect the quality of the reconstruction.

Motivated with recent observations about the effectiveness of neural networks for modeling smooth fields, we propose a neural network-based technique for vector tomography for the purpose of reconstructing a magnetization vector field. When compared against iterative methods, we observe that the quality of reconstruction remains stable under the increasing magnitude of noise. By modeling the vector field using a neural network, we may produce arbitrarily high-resolution reconstructions, at a cost of training time, without significantly increasing memory needed to represent the values of the vector field.

The paper is organized as follows: in Section~\ref{sec:vectorTomography} we briefly set up the notation and briefly review the theory of tomographic reconstruction of vector fields. In Section~\ref{sec:neuralReconstruction} we introduce our neural network-based technique for reconstructing vector fields from the projections and propose different types of architectures when the vector field admits certain symmetries. Finally, in Section~\ref{sec:numRecons} we perform a number of numerical experiments using different techniques to illustrate the stability of the method under increasing magnitudes of noise.

\section{Vector Tomography and ray transform}\label{sec:vectorTomography}
In this section, we briefly review the theory behind the tomographic reconstruction of vector fields \cite{Sharafutdinov+1994, Sparr1998VectorFT, paternain2013tensor}. Recall that the set of all lines $L\subset \mathbb{R}^n$ can be parametrized by a quotient $\mathcal{L} = S^{n-1} \times \mathbb{R}^n / \sim $ where $(\theta, x)\sim (\theta, x')$ iff $x-x' = \lambda\theta$ for some constant $\lambda\in\mathbb{R}$. Note that $\mathcal{L}\simeq TS^{n-1}$. To each $(\theta,x)\in \mathcal{L}$, we associate a line $L_{(\theta,x)}$ defined as:
\begin{gather}
	L_{(\theta,x)} = \{ (x + t \theta \in \mathbb{R}^n~\vert ~ t \in \mathbb{R}\}
\end{gather}
This allows us to define a ray transform on the space of functions $\mathcal{C}_c^\infty(\mathbb{R})$ with compact support as:
\begin{definition}\label{def:rayScalar} A ray-transform on $\mathbb{R}^n$ is an operator:
\begin{gather}
	\mathcal{R}\colon \mathcal{C}_c^\infty(\mathbb{R}^n)\longrightarrow \mathcal{C}_c^\infty (\mathcal{L})
\end{gather}
defined as an integral:
\begin{gather}
	\mathcal{R}(f) 	(L) = \int _L f\,ds,\quad\text{for}~L \in \mathcal{L}
\end{gather}
\end{definition}
A standard result in tomographic reconstruction of scalar functions is that $\mathcal{R}$ is invertible \cite{deans2007radon}.

There is a natural generalization of the ray-transform to the space of functions which are sections of certain bundles, such as tensor fields \cite{Sharafutdinov+1994}. We are mainly interested in the generalization of the ray-transform to the spaces of vector-valued functions. It is easy to see that Def.~\ref{def:rayScalar} naturally generalizes to vector-valued functions, by extending the operator $\mathcal{R}$ linearly. In particular, we have:
\begin{definition} A vector ray-transform for functions on $\mathbb{R}^n$ with values in a vector space $V$ is defined by:
\begin{gather}
	\mathcal{R}(F) = \mathcal{R}\left(\sum_{i=1}^{\dim V}f_i e_i\right) = \sum_{i=1}^{\dim{V}}\mathcal{R}(f_i)e_i
\end{gather}
where $\{e_i\}_i$ is a basis of $V$.
\end{definition}
It is trivial to see that the definition of vector-ray transform is independent from the choice of basis of $V$. Furthermore, the invertibility of the extended operator $\mathcal{R}$ to $V$ follows immediately from the invertibility of the scalar ray-transform. However, note that this requires $\dim{V}$ independent measurements, which poses a significant difficulty in practice, where we are limited to only a subset of the measurements. More generally, we may consider some measurements, which are some linear transformations of the $\mathcal{R}(F)$. In particular, we may consider a probed measurement \cite{Prince1994}:
\begin{definition} A probe is a map $p\colon \mathcal{L}\rightarrow V$. A probed ray-transform of a $V$-valued function $F$ is:
\begin{gather}
	\mathcal{R}_p(F)(L) = \left\langle p(L),~\mathcal{R}(F)(L)\right\rangle
\end{gather}
where $\langle-,-\rangle$ is an inner product on $V$. For a set of $k$ different probes $\mathcal{P} = \{p_1,\dots, p_k\}$, we may form a set of probed measurements:
\begin{gather}
	\mathcal{R}_\mathcal{P}(F) := \left(\mathcal{R}_{p_1}(F),\mathcal{R}_{p_2}(F),\dots, \mathcal{R}_{p_k}(F)\right)	
\end{gather}
or, alternatively $\mathcal{R}_\mathcal{P}(F)(L) = A_\mathcal{P}(L)\cdot \mathcal{R}(F)(L)$ where $A_{\mathcal{P}}(L)$ is a matrix given by:
\begin{gather}
	A_\mathcal{P}(L) = \begin{pmatrix}
 	p_1(L)\\ p_2(L) \\ \vdots \\ p_k(L)
 \end{pmatrix}	
\end{gather}

\end{definition}
For general $V$-valued functions $F$, in order to guarantee the uniqueness of the reconstruction, the matrix $A(L)$ must be of maximal rank for every $L\in \mathcal{L}$ \cite{Prince1994}. However, in the case when the set of probes is limited, we may still guarantee uniqueness on the certain quotients of the spaces of $V$-valued functions, when $V = \mathbb{R}^n$. In particular, consider a single probe $p$:
\begin{gather}\label{eq:probe}
	p(L_{(\theta, x)}) = \tau_L,
\end{gather}
where $\tau_L$ is the unit vector along $L$. It is easy to see that the probed measurement $\mathcal{R}_p(F)(L)$ corresponds to an integral of the component of $F$ along $L$:
\begin{gather}\label{eq:lineIntegral}
	\mathcal{R}_p(F)(L) = \int_L F\cdot dr
\end{gather}
If only a single probe is considered, the probed ray-transform $\mathcal{R}_p$ is not one-to-one. This lack of uniqueness is captured by the following theorem \cite{doi:10.1137/20M1344779}:
\begin{thm}\label{thm:Kernel} The mapping $\mathcal{R}_p$ on	 the space of vector fields has kernel:
\begin{gather}
	\ker\mathcal{R}_p \simeq \mathrm{im}~\nabla~,
\end{gather}
where $\nabla$ is the usual gradient operator on vector fields.
\end{thm}
Therefore, if $\tilde{F}$ is a reconstructed vector field from the probed measurements, with the probe \eqref{eq:probe}, of a general vector field $F$, then: $\tilde{F} - F = \nabla \phi$ for some scalar $\phi$, i.e. only the solenoidal component of $F$ can be reconstructed.

\section{Neural Reconstruction}\label{sec:neuralReconstruction}

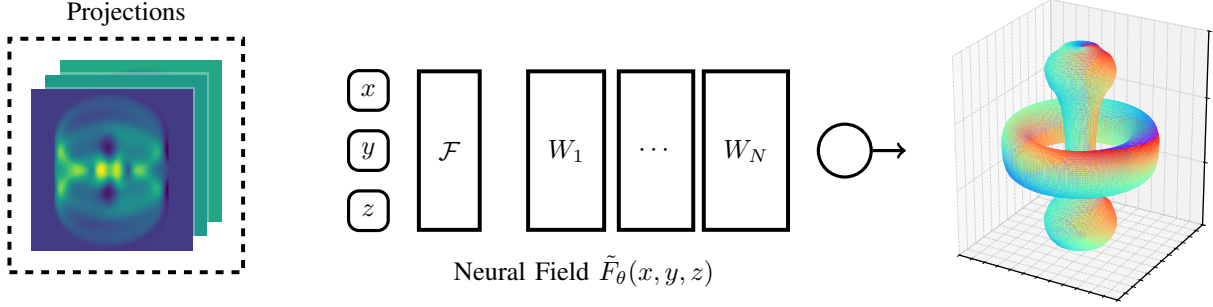
\begin{figure*}
	\centering	
	\begin{tikzpicture}[shorten >= 1pt, ->, draw=black, line width=1.5]
		\tikzstyle{RR}=[very thick, rounded corners, draw=black, minimum size=15];
		\tikzstyle{Split} = [rectangle split, rectangle split parts=2, rectangle split part fill={blue!30,red!20}, rounded corners, draw=black, very thick, minimum size=23, inner sep=5pt, text centered];
		\tikzstyle{neuron}=[very thick,circle,draw=black,minimum size=20,inner sep=0.5,outer sep=0.6]

		\node[opacity=0.3] (img1) at (-0.15cm,-0.045cm,1/2) {\includegraphics[width=2.2cm,height=2.2cm]{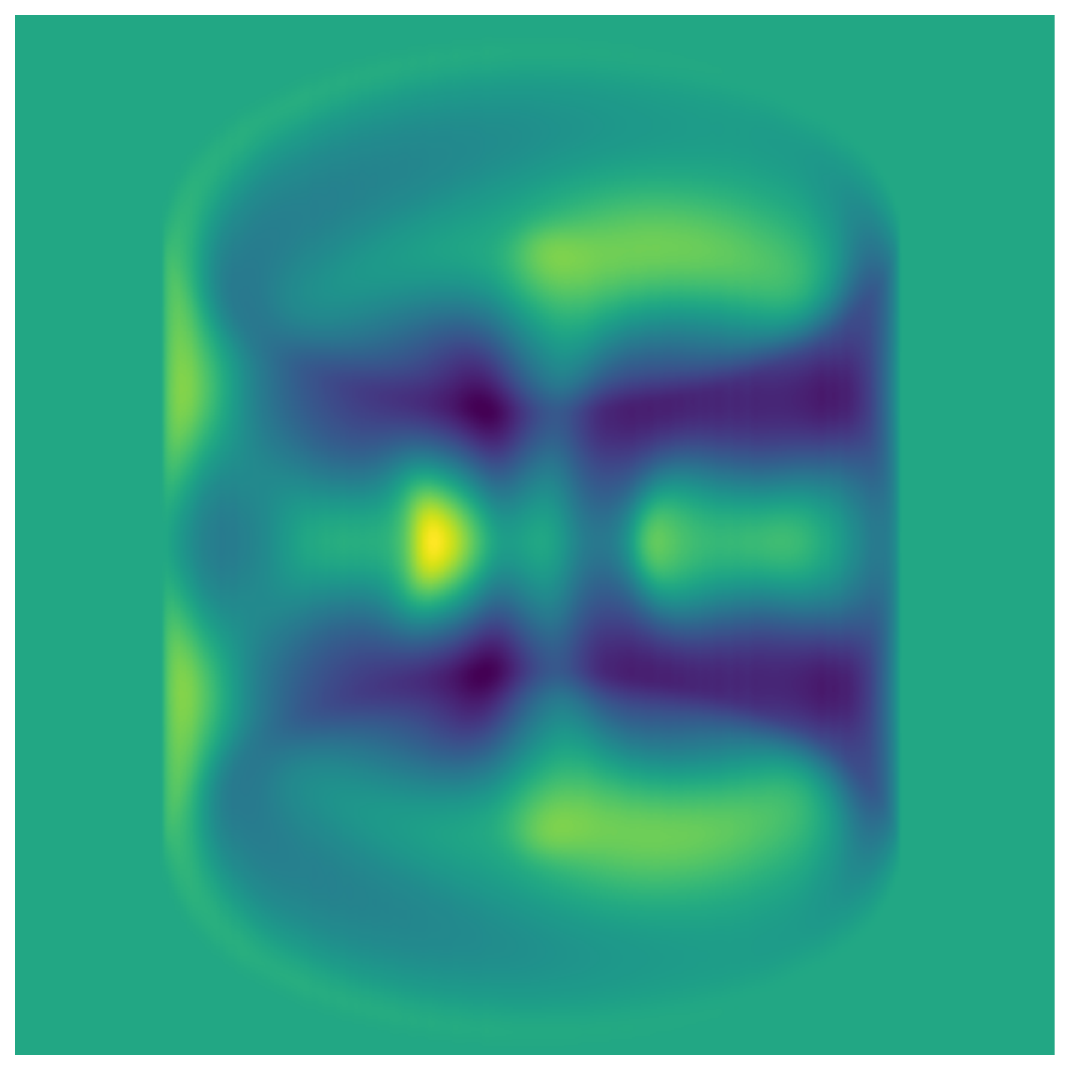}};
		\node[opacity=0.3] (img2) at (-0.15cm,-0.045cm,2/2) {\includegraphics[width=2.2cm,height=2.2cm]{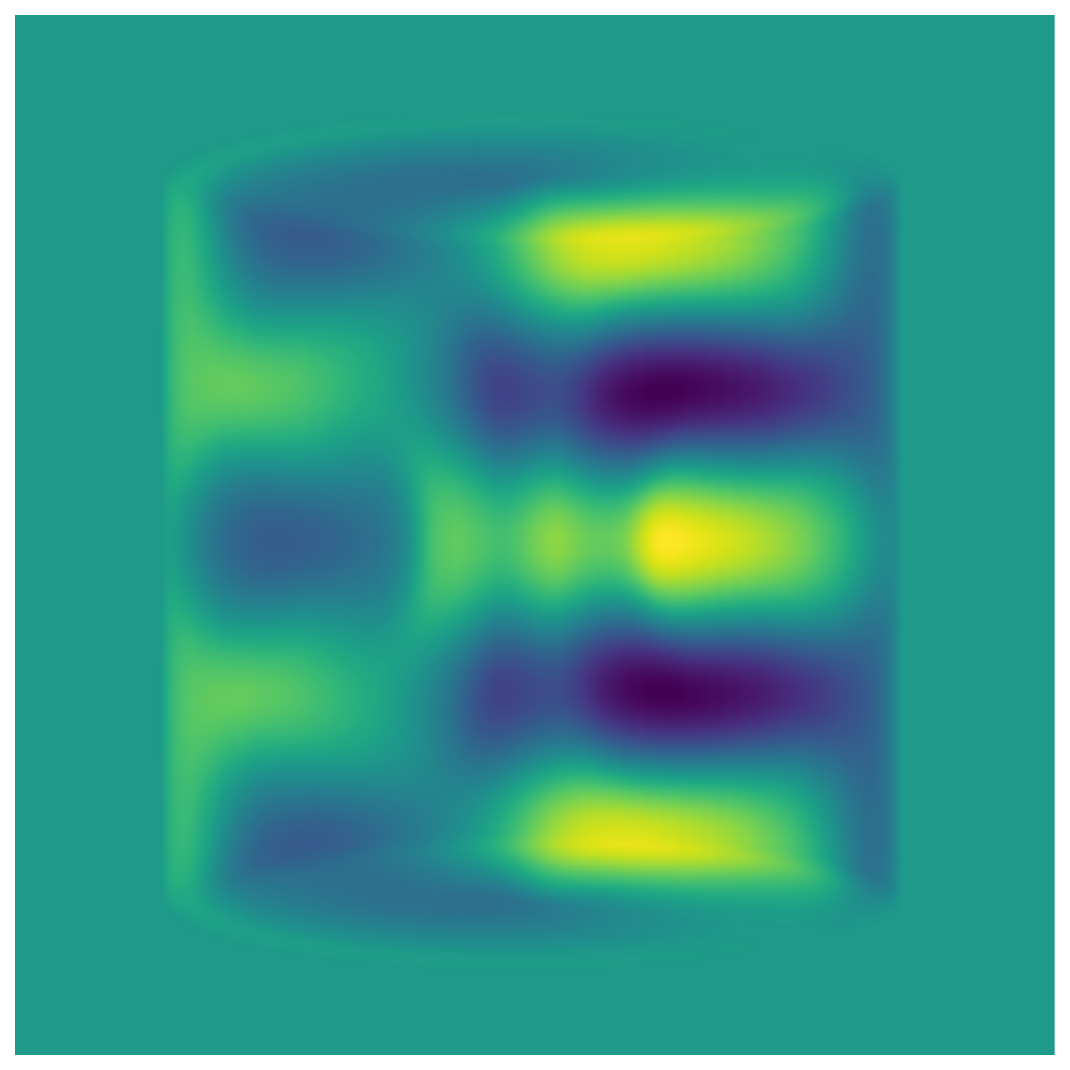}};
		\node[opacity=0.3] (img3) at (-0.15cm,-0.045cm,3/2) {\includegraphics[width=2.2cm,height=2.2cm]{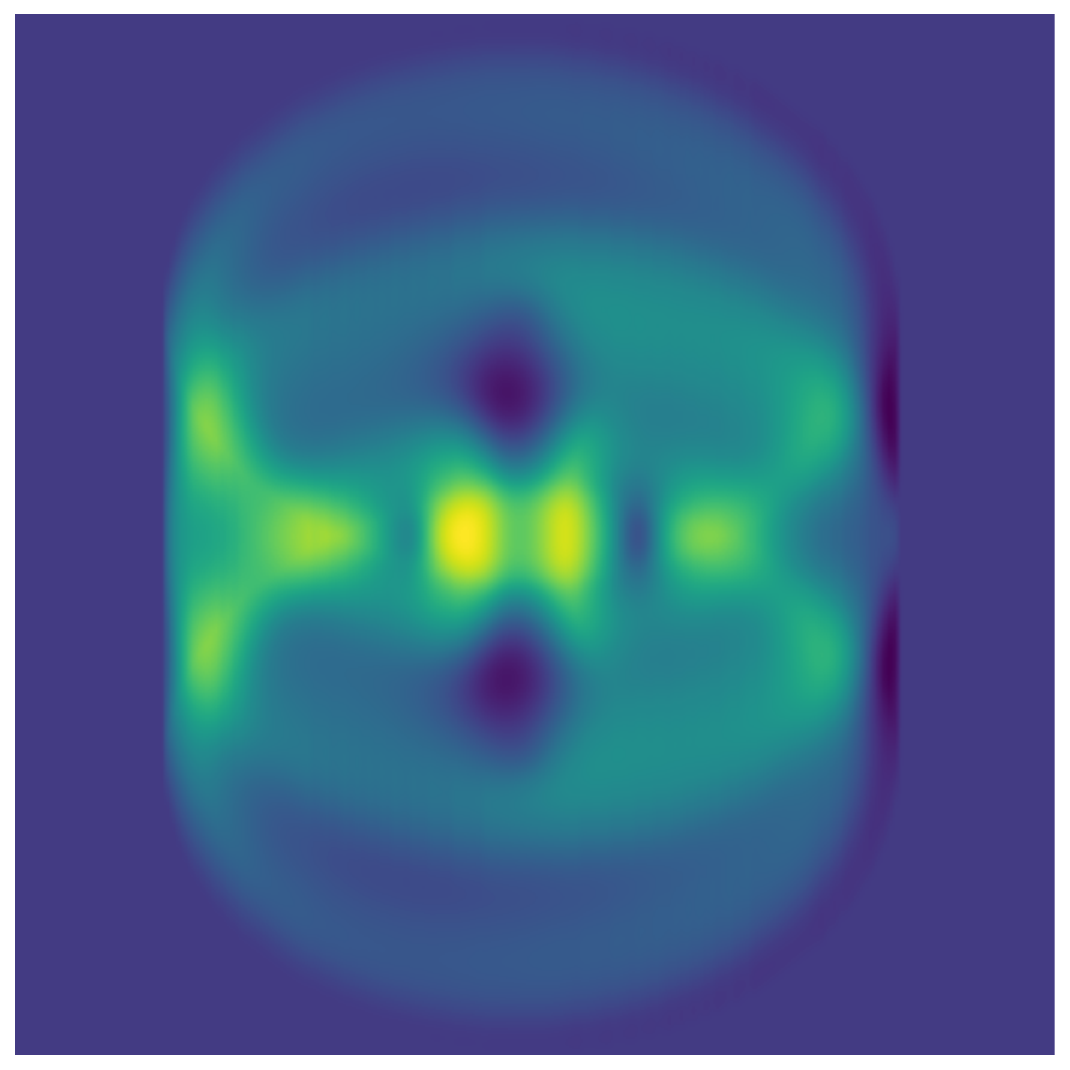}};
	
		\node[draw, dashed, rectangle, minimum height=60pt, minimum width=1pt, fit=(img1) (img2) (img3), label=above:{Projections}] (projectionsBorder) {};

		\foreach \name / \y in {1,...,3}
			\node[RR] (I-\name) at (-1.2*\layersep,-0.8*\y cm) {
				\ifnum \y = 1
					$x$%
				\fi
				\ifnum \y = 2
					$y$%
            	\fi
				\ifnum \y = 3
					$z$%
				\fi};

		\path[yshift=(\widthNN-4)*0.5cm]
			node (F) at (-0.3*\layersep, -0.8*2 cm) {$\mathcal{F}$};
		\node[draw, solid, rectangle, minimum height=60pt, minimum width=1pt, fit=(F)] (FB) {};

		\path[yshift=(\widthNN-4)*0.5cm]
			node (N1) at (\layersep, -0.8*2 cm) {$W_1$};
        \node[draw, solid, rectangle, minimum height=60pt, minimum width=15pt, fit=(N1)]  (NB1) {};

		\path[yshift=(\widthNN-4)*0.5cm]
			node (Ndots) at (2*\layersep, -0.8*2 cm) {$\cdots$};
        \node[draw, solid, rectangle, minimum height=60pt, minimum width=15pt, fit=(Ndots)] (NBdots) {};

		\path[yshift=(\widthNN-4)*0.5cm]
			node (NN) at (3*\layersep, -0.8*2 cm) {$W_N$};
		\node[draw, solid, rectangle, minimum height=60pt, minimum width=10pt, fit=(NN)] (NBN) {};

		\path[yshift=0.0cm]
			node[neuron, pin={[pin edge={->, black, very thick}]right:}] (phi) at (4.1*\layersep, -0.8*2) {};

        \node at (1.2*\layersep, -3.2) {Neural Field $\tilde{F}_{\theta}(x,y,z)$};

		\node[opacity=0.3] (recon) at (6.7*\layersep,-0.8*2cm) {\includegraphics[width=4.5cm,height=4.5cm]{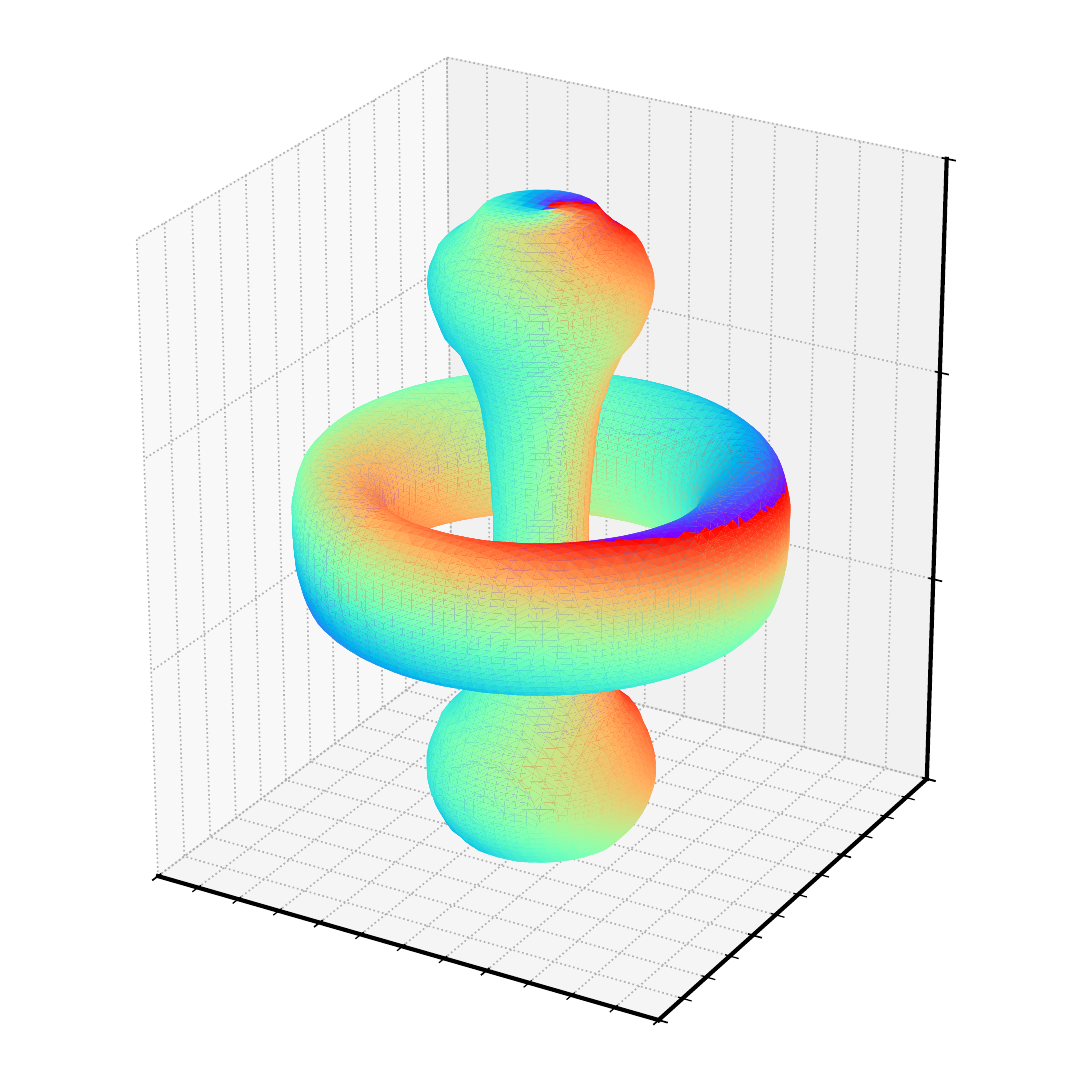}};
	\end{tikzpicture}
	\caption{Outline of the neural field reconstruction of the magnetization vector field.}\label{fig:NN}
\end{figure*}

Let $\mathcal{P}$ be a set of probes on a vector space $V$, and $\mathcal{R}_\mathcal{P}$ be a probed ray-transform. The inverse of the operator $\mathcal{R}_{\mathcal{P}}$ may be approximated by sampling a large number of lines in $\mathcal{L}$ to compute certain integrals over $\mathcal{L}$ \cite{Prince1994,deans2007radon}. However, when the set of lines sampled from $\mathcal{L}$ is limited, these reconstruction techniques may suffer from significant artifacts. This can be circumvented by reformulating the calculation of the inverse as an optimization problem. In particular, let $F\in \mathcal{C}_c^\infty(\mathbb{R}^n,V)$ and consider a set of $N$ probed measurements $\mathcal{R}_{\mathcal{P}}(F)(L_i)$ for $\{L_i\}_{i=1}^N\subset \mathcal{L}$. Then, the problem of recovering $F$ can be formulated as:
\begin{gather}\label{eq:optimization}
	\tilde{F} = \argmin_{\tilde{F}\in S}\frac{1}{N}\sum_{i=1}^N \norm{\mathcal{R}_{\mathcal{P}}(\tilde{F})(L_i) - \mathcal{R}_{\mathcal{P}}(F)(L_i)},
\end{gather}
where $S$ is a space of $V$-valued functions on $\mathbb{R}^n$. Once $S$ is determined, one may then use the usual numerical optimization techniques, such as gradient descent, in order to recover $\tilde{F}$.

In the recent work \cite{Donnelly_2018}, the authors discretize the support of $F$ into voxels $X = \{x_1,\dots , x_l\}\subset \mathbb{R}^n$ and perform point-wise optimization of the values $\tilde{F}(x)$ at each voxel $x\in X$ using gradient descent. Though this method is robust when the number of angles is limited, it requires a large amount of memory to store high-resolution reconstructions. Furthermore, by introducing a noise to the measurements $\mathcal{R}_\mathcal{P}(F)(L_i)$, we observe a significant deterioration of the quality of the reconstruction with the increasing scale of noise.

Motivated with these observation, we instead propose a neural network architecture to approximate the space $S$ of $V$-valued functions on $\mathbb{R}^n$. Similarly to the construction that's used \cite{mildenhall2020nerf, ruckert2022neat} to construct Neural Radiance Fields (NeRFs), we model the field $\tilde{F}$ with a neural network $\tilde{F}_\theta$ (see Fig.~\ref{fig:NN}), where $\theta$ is the set of parameters/weights of the neural network. This reduces the optimization problem \eqref{eq:optimization} into the usual neural network training with the loss function:
\begin{gather}\label{eq:recLoss}
	L_{\mathrm{rec}}(\theta) = \frac{1}{N}\sum_{i=1}^N \norm{\mathcal{R}_\mathcal{P}(\tilde{F}_\theta)(L_i) - \mathcal{R}_\mathcal{P}(F)(L_i)}^2.
\end{gather}
Furthermore, given that this representation does not discretize the domain of $\tilde{F}_\theta$, we may compute the probed measurements $\mathcal{R}_\mathcal{P}(\tilde{F}_\theta)(L_i)$ with arbitrarily high accuracy, by appropriately sampling points along the line $L_i$ without significantly increasing the memory usage. Furthermore, parametrization of the reconstruction $\tilde{F}$ using a neural network allows us to impose additional smoothness conditions on $\tilde{F}_\theta$ to further suppress irregularities induced by noise. In particular, we impose an additional regularization term, on a set of points $\{x_j \sim \mathcal{U}(\supp{F})\}_j$ which are uniformly sampled from the support of $F$:
\begin{gather}\label{eq:regLoss}
	L_{\mathrm{reg}}(\theta) = \sum_{j} \norm{\nabla \tilde{F}_\theta(x_j)}^2,
\end{gather}
and optimize a weighted combination of \eqref{eq:recLoss} and \eqref{eq:regLoss}:
\begin{gather}
	L_\alpha(\theta) = L_{\mathrm{rec}}(\theta) + \alpha L_{\mathrm{reg}}(\theta),
\end{gather}
where $\alpha\geq 0$ controls the strength of smoothing.
\begin{figure*}[htb]
    \centering
	\includegraphics[width=1.0\textwidth]{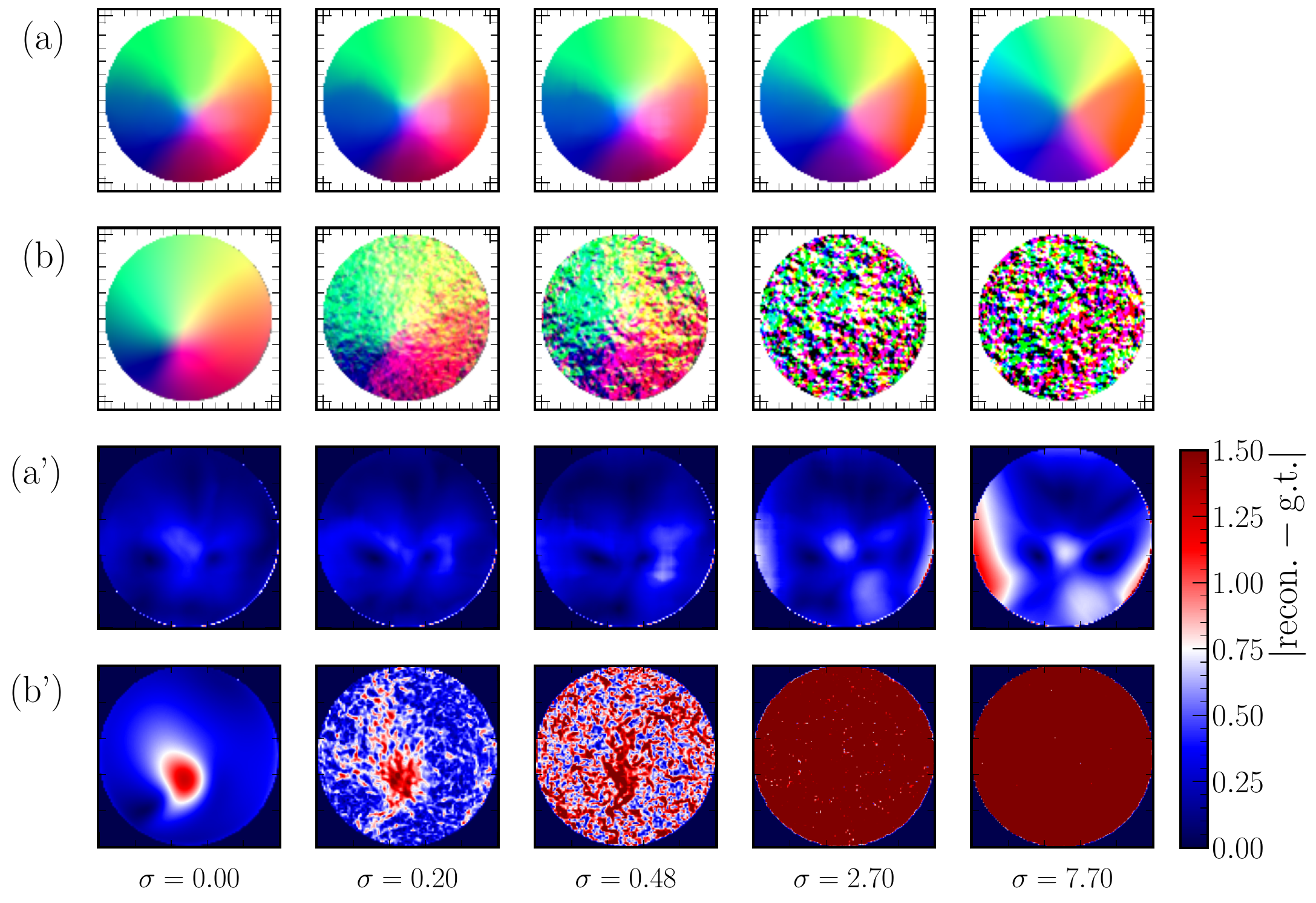}
    \caption{Slices of reconstructions at different levels of noise $\epsilon \sim \mathcal{N}(0,\sigma)$. (a) Reconstruction using neural vector tomography (ours), (b) reconstruction using the method described in \cite{Donnelly_2018} \protect\footnotemark. (a') and (b') show the errors (defined as the absolute value of the difference between reconstruction and the ground truth) for (a) and (b), respectively.}
    \label{fig:comparisonsBloch}
\end{figure*}

\begin{figure}[ht!]
    \centering
    \includegraphics[width=1.\columnwidth]{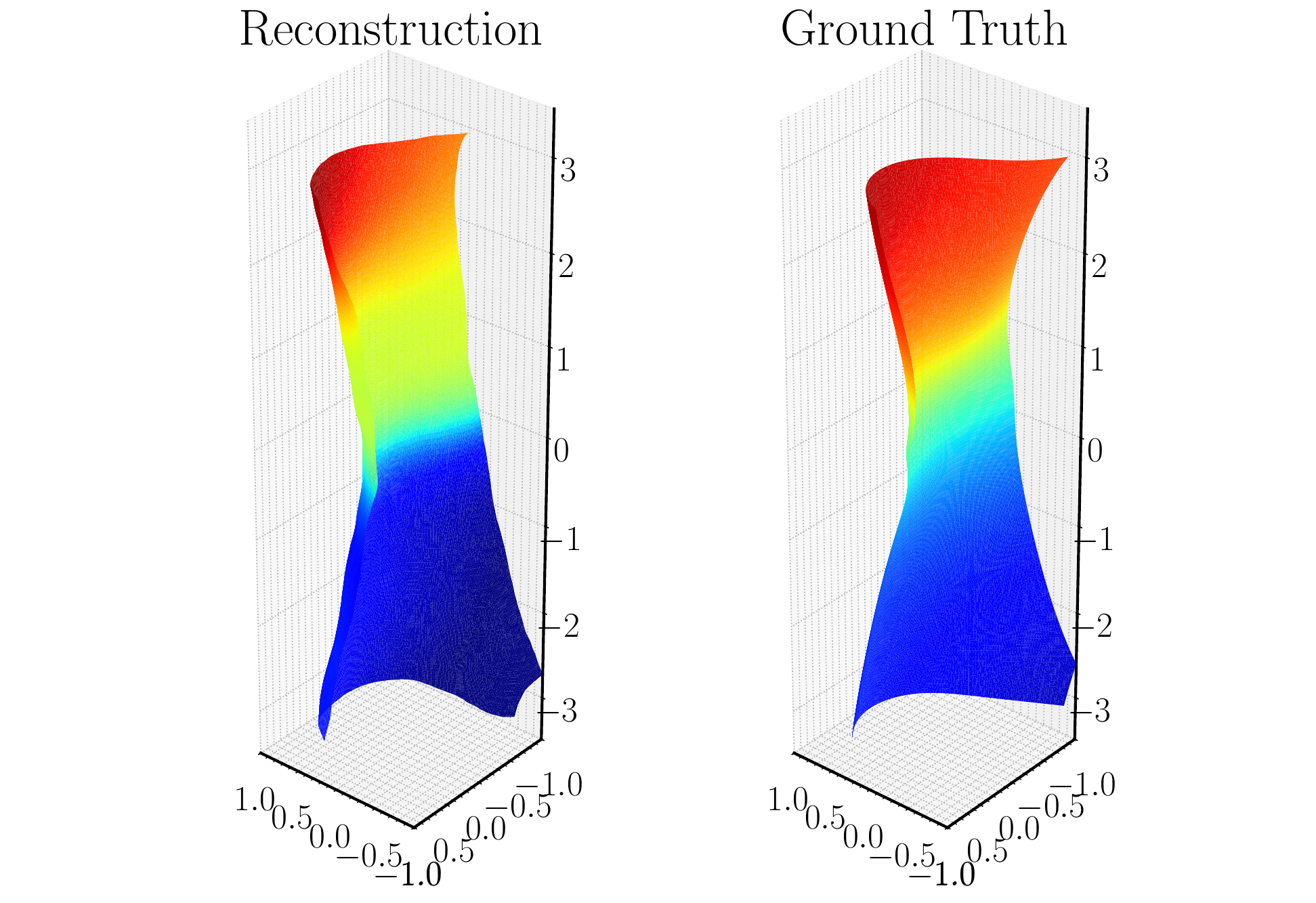}
    \caption{Comparison of the reconstructed and ground truth isosurfaces $M_z = 0$.}\label{fig:bloch3D}
\end{figure}

\begin{figure*}[htb]
    \centering
	\includegraphics[width=1.0\textwidth]{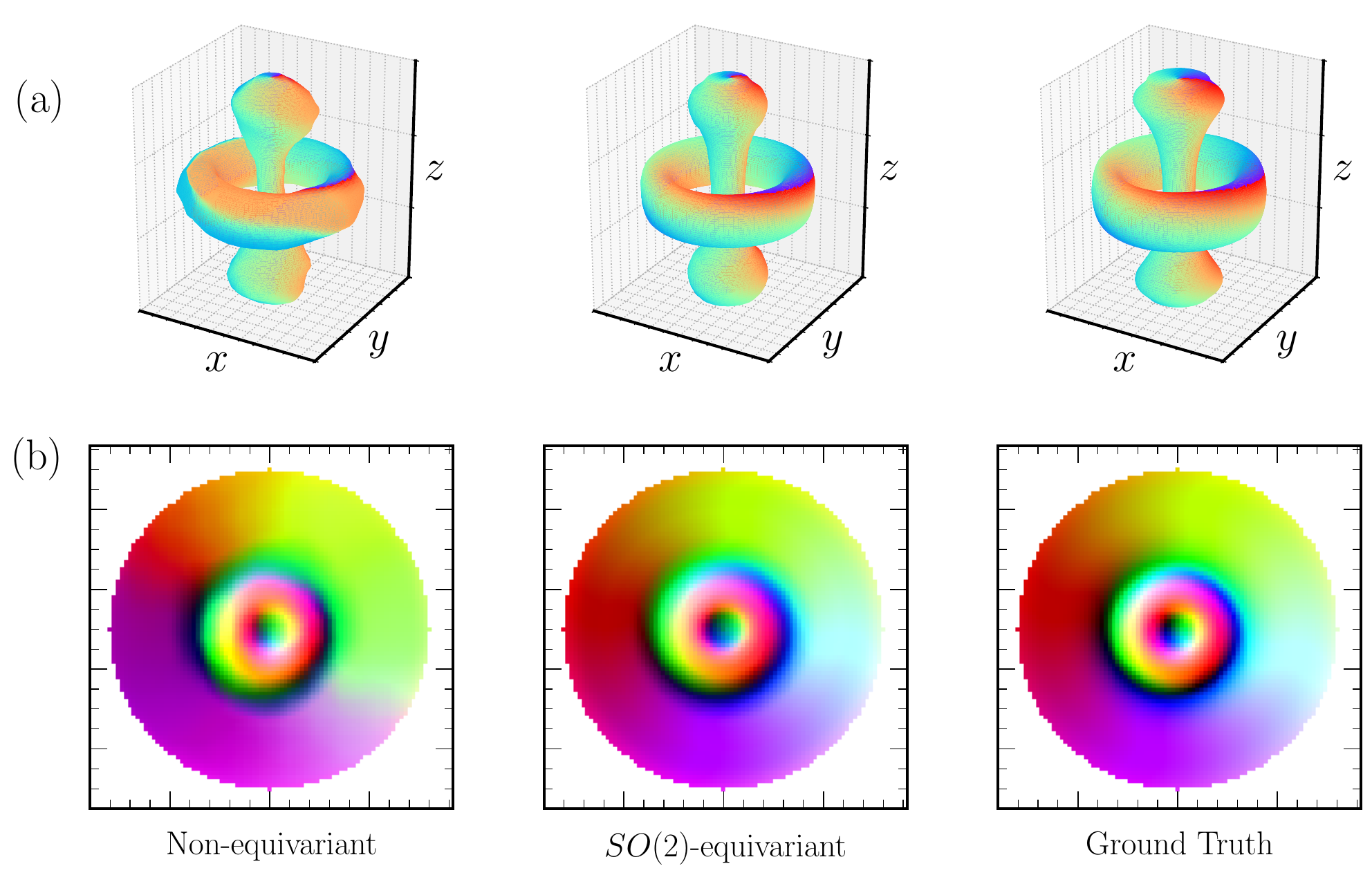}
    \caption{Comparison of (a) isosurfaces $M_z = 0$ and (b) slices of reconstructions using different reconstruction techniques.}
    \label{fig:comparisonsHopfion}
\end{figure*}

\subsection{Sampling Rays}
We are mainly interested in the case when the rays $\{L_i\}_{i=1}^N$ are arranged on a $n$-by-$m$ grid ($N=nm$), so that the measurements $\mathcal{R}_\mathcal{P}(F)(L_i)$ form an image: $M_{i,j} = \mathcal{R}_{\mathcal{P}}(F)(L_{i,j})$ for $1 \leq i \leq n$ and $1 \leq j \leq m$. As the resolution $(n,m)$ of the measurement grows, so does the number of terms in \eqref{eq:recLoss}. Due to limited memory resources, this poses a significant challenge. However, noting that each line $L$ appears as a summand in $L_{\mathrm{rec}}(\theta)$, we may instead randomly subsample constant number of lines $L_{i,j}$ from the full $n$-by-$m$ picture at each training step and replace the full $M_{i,j}$ image with a subsampled rays. This allows us to run reconstructions on measurements with arbitrarily high resolution without exceeding the available memory resources.

\subsection{Approximating $\mathcal{R}_{\mathcal{P}}(\tilde{F}_\theta)(L)$}
In order to optimize \eqref{eq:recLoss}, the approximation of $\mathcal{R}_{\mathcal{P}}(\tilde{F}_\theta)(L)$ must be differentiable w.r.t. the neural network parameters $\theta$. Recall that the ray transform can be computed by computing the following integral:
\begin{gather}
	\mathcal{R}_{p}(\tilde{F}_\theta)(L) = \int_L p(L)\cdot \tilde{F}_{\theta}\,ds
\end{gather}
for every probe $p\in \mathcal{P}$. In order to ensure uniform training of the neural network, we use stratified point sampling along the line $L \cap \supp{F}$, as described in \cite{mildenhall2020nerf}, to approximate the integral over $L_{(\alpha, x)}$ with a probe $p\in \mathcal{P}$ as a sum:
\begin{gather}
	\int_L p(L)\cdot \tilde{F}_{\theta}\,ds  \approx \frac{1}{M}\sum_{i=1}^{M} p(L)\cdot \tilde{F}_{\theta}(x+\alpha t_i)%
\end{gather}
where $t_i \sim \mathcal{U}(I_i)$  where $I_i$ is the $i^\text{th}$ block in the partition of $L\cap \supp{F}$.

\subsection{Symmetries and equivariant neural tomography}
Let $F$ be a vector field admitting a symmetry under the action of a certain group $G$, that is: $g\cdot F = g$ for any $g\in G$. Existence of such symmetry significantly constrains the vector field, and therefore reduces the space $S$ of admissible vector fields in \eqref{eq:optimization}. This, in turn, implies that certain neural network features are redundant and can be removed by modifying features appropriately. In the case when $G$ is a finite-group, this invariance can be enforced by simple group averaging operation:
\begin{gather}
	F \mapsto \sum_{g\in G}g\cdot F	
\end{gather}
This can be generalized to the case when $G$ is a Lie group, thus producing a neural network which is by-construction $G$-equivariant \cite{esteves2020theoretical, cohen2016group, villar2021scalars}. Here, we consider the case when the vector field admits $SO(2)$-equivariance, with the action on vector fields induced by the action on the domain:
\begin{gather}\label{eq:equivCondition}
	(g\cdot F)(x) = gF(g^{-1}x),\quad g\in SO(2),~x\in \mathbb{R}^3
\end{gather}
where $SO(2)$ acts on $\mathbb{R}^3$ by fixing $z$-axis. To construct a neural network satisfying the equivariance condition \eqref{eq:equivCondition}, we define the neural network $F_\theta$ as:
\begin{gather}\label{eq:equivNN}
	F_\theta(x,y,z) = R_{\arctan{(y,x)}}G_\theta(\sqrt{x^2+y^2},z)	
\end{gather}
where $R_{arctan{(y,x)}} \in SO(2)$ is a rotation matrix, given by:
\begin{gather}
	R_\alpha = \begin{pmatrix}
 	\cos(\alpha) & -\sin(\alpha)	 & 0 \\
 	\sin(\alpha) & \cos(\alpha) & 0 \\
 	0 & 0 & 1
 \end{pmatrix}
\end{gather}
It is easy to see that a neural network constructed using \eqref{eq:equivNN} indeed satisfies is indeed $SO(2)$-equivariant.

\section{Numerical Reconstructions}\label{sec:numRecons}

\footnotetext{Only the vector tomography part has been optimized. The scalar part has been replaced with the ground truth mask.}

\begin{figure}[h!]
    \centering
	\includegraphics[width=1.\columnwidth]{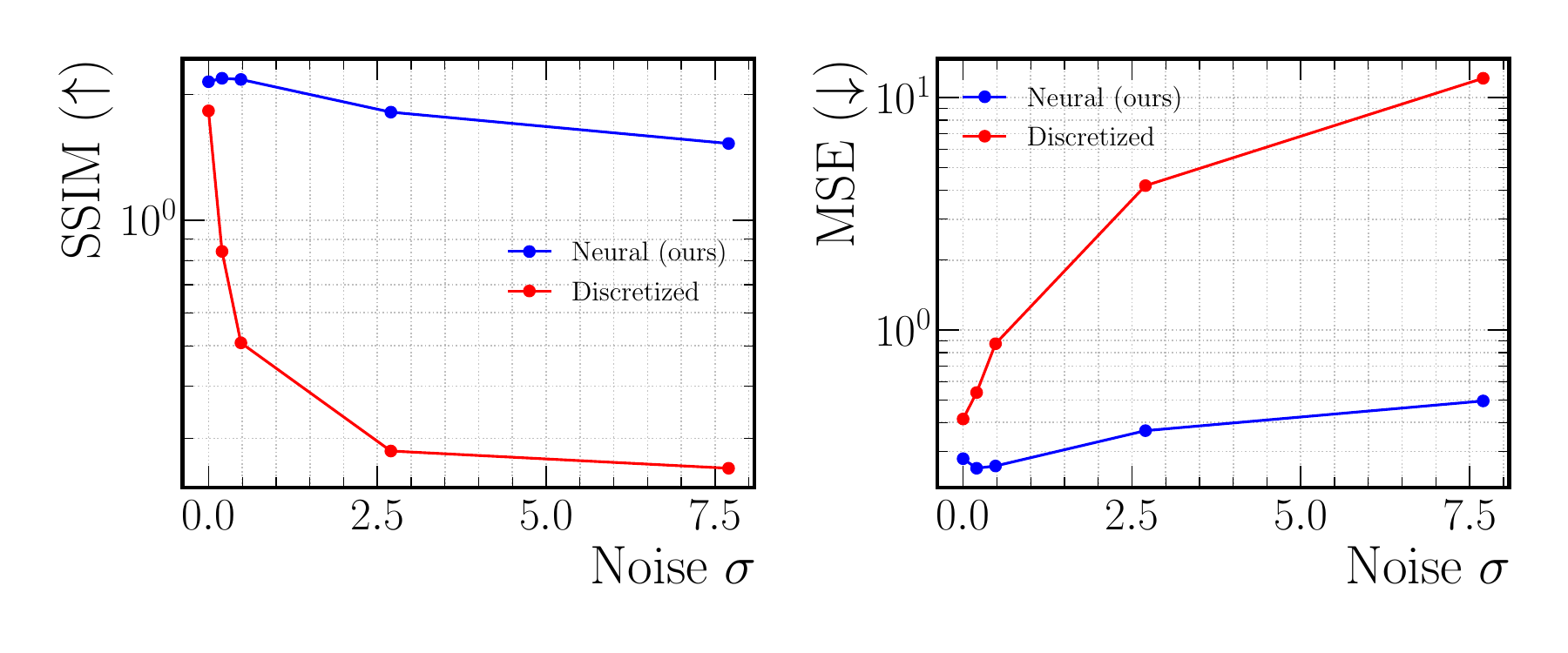}
    \caption{SSIM and MSE metrics of the reconstructions using neural and discrete \cite{Donnelly_2018} techniques under different levels of noise.}
    \label{fig:mseAndSSIMBloch}
\end{figure}

In order to quantify the accuracy of the model, we consider XMCD projections of a simulated magnetization vector field containing two Bloch points \cite{Donnelly_2018}. Recall that X-ray projections correspond to the probed measurements with a probe $p$ given in Eqn.~\eqref{eq:probe}. Therefore, as a consequence of the Thm~\ref{thm:Kernel}, the information pertaining to the exact component of the vector field is lost after the application of the ray-transform operator $\mathcal{R}_p$. However, we may still attempt to reconstruct the vector field, upto exact terms, using the neural reconstruction technique, described in Sec~\ref{sec:neuralReconstruction}.
 We denote the set of rays corresponding to these measurements by $\{L_i\}_{i=1}^N$, and the vector field corresponding to the magnetization by $F$. In order to quantify the robustness of the methods w.r.t. the measurement error, we feed each method with a perturbed set of measurements:
\begin{gather}
	M_i = \mathcal{R}_p(F)(L_i)	 + \epsilon_i
\end{gather}
where $\epsilon_i \sim \mathcal{N}(0,\sigma^2)$ are i.i.d. random variables with the same variance $\sigma$. The comparison of the results of our neural field based and the domain-discretized \cite{Donnelly_2018} techniques is shown on the Fig.~\ref{fig:comparisonsBloch}. Note the stark difference in different methods when the measurements have noise with $\sigma>0$. We observe this to be the general property of neural network-based reconstructions. This can be understood by following observation: in discretized domain, there are no additional conditions favoring a continuous approximation of the vector field. However, in the case when the field is approximated by a neural network, this is implicitly enforced, due to the fact that the activation functions that are used in the neural networks are continuous. The robustness of the method is further supported by comparing the MSE and SSIM scores of the different reconstruction methods, as shown in Fig.~\ref{fig:mseAndSSIMBloch}.

The reconstructions depicted in Fig.~\ref{fig:comparisonsBloch} are produced by a total of 360 rays: 180 angles per each tilt angle, for the total of two tilt axes. The $M_y=0$ isosurface is shown in Figure~\ref{fig:bloch3D}.

For a setup admitting $SO(2)$ symmetry, we consider magnetization vector field induced by a Hopfion \cite{Zheng2023, Lyu2024}. The comparison of the reconstructions produced by equivariant and non-equivariant neural networks are shown on Fig.~\ref{fig:comparisonsHopfion}. The comparison of SSIM and MSE scores is shown in Table~\ref{tab:Comparison}.

\begin{table}[htp]
\centering

\begin{tabular}{ccc}
\toprule
Model & \multicolumn{2}{c}{Metric} \\

\cmidrule(lr){2-3}
& {MSE ($\downarrow$)} & {SSIM ($\uparrow$)} \\
\midrule
Non-equivariant & 0.39 & 2.38 \\
$SO(2)$-equivariant & \B 0.14 & \B 2.79 \\
\bottomrule
\end{tabular}

\caption{SSIM and MSE scores of reconstructions using equivariant and non-equivariant neural networks.}\label{tab:Comparison}

\end{table}

\section{Conclusion}
Noise in the measurements of the magnetization vector fields posits a significant challenge in producing accurate reconstructions of the field.  Methods which model the field over a fixed discretized domain may suffer from significant artifacts when the measurements are imperfect. Motivated with this observation, in this work, we described a neural-network based technique, where the field is described by a neural field. This has allowed to circumvent the limitations posed by the discretization, as the numerical integrals are approximated by values of the field at a set of points that are randomly generated at each iteration. In particular, we have demonstrated the robustness of the technique by artificially introducing increasing levels of noise to the projections. This work was supported by the Office of Basic Energy Sciences, Division of Materials Sciences and Engineering, U.S. Department of Energy under Grant No. DE-SC0020221.
\bibliographystyle{IEEEtran}
\bibliography{ref}

\begin{thebibliography}{10}
\providecommand{\url}[1]{#1}
\csname url@samestyle\endcsname
\providecommand{\newblock}{\relax}
\providecommand{\bibinfo}[2]{#2}
\providecommand{\BIBentrySTDinterwordspacing}{\spaceskip=0pt\relax}
\providecommand{\BIBentryALTinterwordstretchfactor}{4}
\providecommand{\BIBentryALTinterwordspacing}{\spaceskip=\fontdimen2\font plus
\BIBentryALTinterwordstretchfactor\fontdimen3\font minus \fontdimen4\font\relax}
\providecommand{\BIBforeignlanguage}[2]{{%
\expandafter\ifx\csname l@#1\endcsname\relax
\typeout{** WARNING: IEEEtran.bst: No hyphenation pattern has been}%
\typeout{** loaded for the language `#1'. Using the pattern for}%
\typeout{** the default language instead.}%
\else
\language=\csname l@#1\endcsname
\fi
#2}}
\providecommand{\BIBdecl}{\relax}
\BIBdecl

\bibitem{fernandez-pacheco2017}
\BIBentryALTinterwordspacing
A.~Fernández-Pacheco, R.~Streubel, O.~Fruchart, R.~Hertel, P.~Fischer, and R.~P. Cowburn, ``\BIBforeignlanguage{en}{Three-dimensional nanomagnetism},'' \emph{\BIBforeignlanguage{en}{Nature Communications}}, vol.~8, no.~1, p. 15756, Jun. 2017, publisher: Nature Publishing Group. [Online]. Available: \url{https://www.nature.com/articles/ncomms15756}
\BIBentrySTDinterwordspacing

\bibitem{Donnelly_2018}
\BIBentryALTinterwordspacing
C.~Donnelly, S.~Gliga, V.~Scagnoli, M.~Holler, J.~Raabe, L.~J. Heyderman, and M.~Guizar-Sicairos, ``Tomographic reconstruction of a three-dimensional magnetization vector field,'' \emph{New Journal of Physics}, vol.~20, no.~8, p. 083009, aug 2018. [Online]. Available: \url{https://dx.doi.org/10.1088/1367-2630/aad35a}
\BIBentrySTDinterwordspacing

\bibitem{donnelly2020}
\BIBentryALTinterwordspacing
C.~Donnelly and V.~Scagnoli, ``\BIBforeignlanguage{en}{Imaging three-dimensional magnetic systems with x-rays},'' \emph{\BIBforeignlanguage{en}{Journal of Physics: Condensed Matter}}, vol.~32, no.~21, p. 213001, Feb. 2020, publisher: IOP Publishing. [Online]. Available: \url{https://dx.doi.org/10.1088/1361-648X/ab5e3c}
\BIBentrySTDinterwordspacing

\bibitem{hierro-rodriguez2020}
\BIBentryALTinterwordspacing
A.~Hierro-Rodriguez, C.~Quirós, A.~Sorrentino, L.~M. Alvarez-Prado, J.~I. Martín, J.~M. Alameda, S.~McVitie, E.~Pereiro, M.~Vélez, and S.~Ferrer, ``\BIBforeignlanguage{en}{Revealing {3D} magnetization of thin films with soft {X}-ray tomography: magnetic singularities and topological charges},'' \emph{\BIBforeignlanguage{en}{Nature Communications}}, vol.~11, no.~1, p. 6382, Dec. 2020, publisher: Nature Publishing Group. [Online]. Available: \url{https://www.nature.com/articles/s41467-020-20119-x}
\BIBentrySTDinterwordspacing

\bibitem{phatak2010}
\BIBentryALTinterwordspacing
C.~Phatak, A.~K. Petford-Long, and M.~De~Graef, ``Three-{{Dimensional Study}} of the {{Vector Potential}} of {{Magnetic Structures}},'' vol. 104, no.~25, p. 253901. [Online]. Available: \url{https://link.aps.org/doi/10.1103/PhysRevLett.104.253901}
\BIBentrySTDinterwordspacing

\bibitem{tanigaki2015}
\BIBentryALTinterwordspacing
T.~Tanigaki, Y.~Takahashi, T.~Shimakura, T.~Akashi, R.~Tsuneta, A.~Sugawara, and D.~Shindo, ``Three-dimensional observation of magnetic vortex cores in stacked ferromagnetic discs,'' vol.~15, no.~2, pp. 1309--1314. [Online]. Available: \url{https://doi.org/10.1021/nl504473a}
\BIBentrySTDinterwordspacing

\bibitem{wolf2019}
\BIBentryALTinterwordspacing
D.~Wolf, N.~Biziere, S.~Sturm, D.~Reyes, T.~Wade, T.~Niermann, J.~Krehl, B.~Warot-Fonrose, B.~Büchner, E.~Snoeck, C.~Gatel, and A.~Lubk, ``Holographic vector field electron tomography of three-dimensional nanomagnets,'' vol.~2, no.~1, pp. 1--9. [Online]. Available: \url{https://www.nature.com/articles/s42005-019-0187-8}
\BIBentrySTDinterwordspacing

\bibitem{wolf2022}
\BIBentryALTinterwordspacing
A.~Lubk, ``Unveiling the three-dimensional magnetic texture of skyrmion tubes,'' vol.~17, no.~3, pp. 250--255. [Online]. Available: \url{https://www.nature.com/articles/s41565-021-01031-x}
\BIBentrySTDinterwordspacing

\bibitem{lai1994}
\BIBentryALTinterwordspacing
G.~Lai, T.~Hirayama, A.~Fukuhara, K.~Ishizuka, T.~Tanji, and A.~Tonomura, ``Three‐dimensional reconstruction of magnetic vector fields using electron‐holographic interferometry,'' vol.~75, no.~9, pp. 4593--4598. [Online]. Available: \url{https://doi.org/10.1063/1.355955}
\BIBentrySTDinterwordspacing

\bibitem{phatak2008}
\BIBentryALTinterwordspacing
C.~Phatak, M.~Beleggia, and M.~De~Graef, ``Vector field electron tomography of magnetic materials: {{Theoretical}} development,'' vol. 108, no.~6, pp. 503--513. [Online]. Available: \url{https://www.sciencedirect.com/science/article/pii/S0304399107001957}
\BIBentrySTDinterwordspacing

\bibitem{phatak2015}
\BIBentryALTinterwordspacing
C.~Phatak and D.~Gürsoy, ``Iterative reconstruction of magnetic induction using {{Lorentz}} transmission electron tomography,'' vol. 150, pp. 54--64. [Online]. Available: \url{https://www.sciencedirect.com/science/article/pii/S0304399114002496}
\BIBentrySTDinterwordspacing

\bibitem{prabhat2017}
\BIBentryALTinterwordspacing
K.~C. Prabhat, K.~Aditya~Mohan, C.~Phatak, C.~Bouman, and M.~De~Graef, ``{{3D}} reconstruction of the magnetic vector potential using model based iterative reconstruction,'' vol. 182, pp. 131--144. [Online]. Available: \url{https://www.sciencedirect.com/science/article/pii/S0304399117301948}
\BIBentrySTDinterwordspacing

\bibitem{Lyu2024}
\BIBentryALTinterwordspacing
B.~Lyu, S.~Zhao, Y.~Zhang, W.~Wang, F.~Zheng, R.~E. Dunin-Borkowski, J.~Zang, and H.~Du, ``Three-dimensional magnetization reconstruction from electron optical phase images with physical constraints,'' \emph{Science China Physics, Mechanics \& Astronomy}, vol.~67, no.~11, Sep. 2024. [Online]. Available: \url{http://dx.doi.org/10.1007/s11433-024-2448-6}
\BIBentrySTDinterwordspacing

\bibitem{ruckert2022neat}
D.~R{\"u}ckert, Y.~Wang, R.~Li, R.~Idoughi, and W.~Heidrich, ``Neat: Neural adaptive tomography,'' \emph{ACM Transactions on Graphics (TOG)}, vol.~41, no.~4, pp. 1--13, 2022.

\bibitem{mildenhall2020nerf}
B.~Mildenhall, P.~P. Srinivasan, M.~Tancik, J.~T. Barron, R.~Ramamoorthi, and R.~Ng, ``Nerf: Representing scenes as neural radiance fields for view synthesis,'' in \emph{ECCV}, 2020.

\bibitem{lyu2022}
\BIBentryALTinterwordspacing
B.~Lyu, S.~Zhao, Y.~Zhang, W.~Wang, H.~Du, and J.~Zang. {{MagNet}}: Machine learning enhanced three-dimensional magnetic reconstruction. [Online]. Available: \url{http://arxiv.org/abs/2210.03066}
\BIBentrySTDinterwordspacing

\bibitem{Chang2024}
\BIBentryALTinterwordspacing
H.~Chang, V.~Kobzarenko, and D.~Mitra, ``Inverse radon transform with deep learning: an application in cardiac motion correction,'' \emph{Physics in Medicine \& Biology}, vol.~69, no.~3, p. 035010, Jan. 2024. [Online]. Available: \url{http://dx.doi.org/10.1088/1361-6560/ad0eb5}
\BIBentrySTDinterwordspacing

\bibitem{he2020radon}
J.~He, Y.~Wang, and J.~Ma, ``Radon inversion via deep learning,'' \emph{IEEE transactions on medical imaging}, vol.~39, no.~6, pp. 2076--2087, 2020.

\bibitem{Mishra-Sharma:2022bco}
S.~Mishra-Sharma and G.~Yang, ``{Strong Lensing Source Reconstruction Using Continuous Neural Fields},'' in \emph{{39th International Conference on Machine Learning Conference}}, 6 2022.

\bibitem{levis2022gravitationally}
A.~Levis, P.~P. Srinivasan, A.~A. Chael, R.~Ng, and K.~L. Bouman, ``Gravitationally lensed black hole emission tomography,'' in \emph{Proceedings of the IEEE/CVF Conference on Computer Vision and Pattern Recognition}, 2022, pp. 19\,841--19\,850.

\bibitem{Sharafutdinov+1994}
\BIBentryALTinterwordspacing
V.~A. Sharafutdinov, \emph{Integral Geometry of Tensor Fields}.\hskip 1em plus 0.5em minus 0.4em\relax Berlin, New York: De Gruyter, 1994. [Online]. Available: \url{https://doi.org/10.1515/9783110900095}
\BIBentrySTDinterwordspacing

\bibitem{Sparr1998VectorFT}
\BIBentryALTinterwordspacing
G.~Sparr, K.~Str, and A.~En, ``Vector field tomography, an overview,'' 1998. [Online]. Available: \url{https://api.semanticscholar.org/CorpusID:16235452}
\BIBentrySTDinterwordspacing

\bibitem{paternain2013tensor}
G.~P. Paternain, M.~Salo, and G.~Uhlmann, ``Tensor tomography: progress and challenges,'' \emph{arXiv preprint arXiv:1303.6114}, 2013.

\bibitem{deans2007radon}
\BIBentryALTinterwordspacing
S.~Deans, \emph{The Radon Transform and Some of Its Applications}, ser. Dover Books on Mathematics Series.\hskip 1em plus 0.5em minus 0.4em\relax Dover Publications, 2007. [Online]. Available: \url{https://books.google.com/books?id=xSCc0KGi0u0C}
\BIBentrySTDinterwordspacing

\bibitem{Prince1994}
\BIBentryALTinterwordspacing
J.~Prince, ``Tomographic reconstruction of 3-d vector fields using inner product probes,'' \emph{IEEE Transactions on Image Processing}, vol.~3, no.~2, p. 216–219, Mar. 1994. [Online]. Available: \url{http://dx.doi.org/10.1109/83.277903}
\BIBentrySTDinterwordspacing

\bibitem{doi:10.1137/20M1344779}
\BIBentryALTinterwordspacing
J.~Ilmavirta and K.~M\"{o}nkk\"{o}nen, ``X-ray tomography of one-forms with partial data,'' \emph{SIAM Journal on Mathematical Analysis}, vol.~53, no.~3, pp. 3002--3015, 2021. [Online]. Available: \url{https://doi.org/10.1137/20M1344779}
\BIBentrySTDinterwordspacing

\bibitem{esteves2020theoretical}
C.~Esteves, ``Theoretical aspects of group equivariant neural networks,'' \emph{arXiv preprint arXiv:2004.05154}, 2020.

\bibitem{cohen2016group}
T.~Cohen and M.~Welling, ``Group equivariant convolutional networks,'' in \emph{International conference on machine learning}.\hskip 1em plus 0.5em minus 0.4em\relax PMLR, 2016, pp. 2990--2999.

\bibitem{villar2021scalars}
S.~Villar, D.~W. Hogg, K.~Storey-Fisher, W.~Yao, and B.~Blum-Smith, ``Scalars are universal: Equivariant machine learning, structured like classical physics,'' \emph{Advances in Neural Information Processing Systems}, vol.~34, pp. 28\,848--28\,863, 2021.

\bibitem{Zheng2023}
\BIBentryALTinterwordspacing
F.~Zheng, N.~S. Kiselev, F.~N. Rybakov, L.~Yang, W.~Shi, S.~Bl\"{u}gel, and R.~E. Dunin-Borkowski, ``Hopfion rings in a cubic chiral magnet,'' \emph{Nature}, vol. 623, no. 7988, p. 718–723, Nov. 2023. [Online]. Available: \url{http://dx.doi.org/10.1038/s41586-023-06658-5}
\BIBentrySTDinterwordspacing

\end{thebibliography}

\end{document}